\documentstyle[ltwol]{article}

\begin{document}

\title{EXPERIMENTAL IMPLICATIONS OF THE DUAL COLOUR SOLUTION TO THE 
   GENERATION PUZZLE}

\author{CHAN HONG-MO}

\address{Rutherford Appleton Laboratory,\\
  Chilton, Didcot, Oxon, OX11 0QX, United Kingdom\\E-mail:
chanhm\,@\,v2.rl.ac.uk}

\twocolumn[\maketitle\abstracts{
Apart from offering explanations not only for the distinctive fermion mass 
and mixing patterns but also for the actual values of the mass and mixing 
parameters, the dual colour solution to the generation puzzle gives numerous 
detailed predictions ranging from rare FCNC meson decays and $\mu$-$e$ 
conversions in nuclei at low energies to cosmic ray air showers with energies 
beyond $10^{20}$ eV at the extreme end of the experimental range.  Besides, 
it predicts a new class of flavour-violating phenomena (called transmutations) 
due to the rotating fermion mass matrix which are unambiguously calculable. 
Comparison with experiment of these many ``parameter-free'' predictions 
reveals no violation of existing bounds but identifies several striking
effects which can be tested with present experimental sensitivity.}]

The proposed dual colour solution to the generation puzzle as contained in 
the so-called Dualized Standard Model (DSM) scheme offers not only a {\it 
raison d'\^etre} for 3, and apparently only 3, generations of fermions, but 
also an explanation for their distinctive mass and mixing patterns together
with near-quantitative calculations for the values of their mass and mixing 
parameters~\cite{dsmrevs}.  A very brief summary of these results and of the 
theoretical framework on which they are based is given in the talk by my 
collaborator Tsou in this conference~\cite{Tsou} to which the present talk
may be regarded as a sequel.  As an example for the quality of the overall 
agreement with experiment in these respects, we quote here in Table 
\ref{CKMtable} from one of our papers~\cite{phenodsm} the results on the
quark and lepton mixing parameters and compare them with the latest 
experimental values~\cite{databook}.  All the predictions shown were 
obtained by adjusting the only remaining free parameter in the scheme 
to fit the Cabibbo angle, the other parameters having already been fixed 
earlier by fitting the masses of quarks and leptons.  One sees in the 
table that the predicted range overlaps more or less with the experimental 
range in all cases except for the element $U_{e2}$ in the lepton mixing 
(MNS) matrix relevant for the oscillations of solar neutrinos which is 
known to be particularly difficult for the DSM scheme to calculate.

\begin{table}[ht]
\begin{eqnarray*}
\begin{array}{||c||c||c||}  
\hline \hline
         & {\rm Experimental} & {\rm Predicted} \\
         & {\rm Range}        & {\rm Range}     \\
\hline \hline
|V_{ud}| & 0.9742 - 0.9757 & 0.9745 - 0.9762 \\ \hline
|V_{us}| & 0.219 - 0.226 & input \\ \hline
|V_{ub}| & 0.002 - 0.005 & 0.0043 - 0.0046 \\ \hline
|V_{cd}| & 0.219 - 0.225 & input \\ \hline
|V_{cs}| & 0.9734 - 0.9749 & 0.9733 - 0.9756 \\ \hline
|V_{cb}| & 0.037 - 0.043 & 0.0354 - 0.0508 \\ \hline
|V_{td}| & 0.004 - 0.014 & 0.0120 - 0.0157 \\ \hline
|V_{ts}| & 0.035 - 0.043 & 0.0336 - 0.0486 \\ \hline
|V_{tb}| & 0.9990 - 0.9993 & 0.9988 - 0.9994 \\ \hline
|V_{ub}/V_{cb}| & 0.08 \pm 0.02 & 0.0859 - 0.1266 \\ \hline
|V_{td}/V_{ts}| & < 0.27 & 0.3149 - 0.3668 \\ \hline
|V_{tb}^{*}V_{td}| & 0.0084 \pm 0.0018 & 0.0120 - 0.0156 \\ \hline
   \hline
|U_{\mu3}| & 0.56 - 0.83 & 0.6528 - 0.6770 \\ \hline
|U_{e3}| & 0.00 - 0.15 & 0.0632 - 0.0730 \\ \hline
|U_{e2}| & 0.4 - 0.7 & 0.2042 - 0.2531 \\ \hline \hline 
\end{array}
\end{eqnarray*}
\caption{Predicted CKM matrix elements for both quarks and leptons}
\label{CKMtable}
\end{table} 

Any new assumptions introduced by the scheme being by now quite tightly 
constrained by the above-cited calculation of fermion mass and mixing
parameters results in a highly predictive framework with wide-ranging 
ramifications.  A brief resum\'e of the consequences so far examined and 
of their comparison with experiment is the purpose of the present paper.

Two broad areas of application of the scheme arise as follows.  

First, since in the DSM scheme generation is identified with dual colour 
which, like colour, is a local gauge symmetry, any particles carrying the 
generation index can interact via the exchange of dual colour bosons.  This
new interaction can change the generation index and will thus manifest itself
most clearly in what are called flavour-changing neutral current (FCNC) 
effects.  The size of the effects so induced depends on the mass scale
of the dual colour bosons, which is unfortunately not constrained by the
fit to the fermion mass and mixing patterns cited above~\cite{dsmrevs}.
It has thus to be taken as a new parameter, which is presumably quite large, 
or otherwise the new bosons would have already been seen.  With just this 
one free parameter, however, DSM is then able to predict most FCNC effects 
in detail.  The reason is that, in contrast to most other FCNC models, the
DSM scheme with its particular symmetry-breaking mechanism for generations
suggested by duality, is constrained by the empirical fermion mass and
mixing patterns to give specific predictions also for the (perturbative) 
mass spectrum of the dual colour bosons as well as for the branching 
coefficients of their coupling to the various fermion states, while the 
coupling strength of dual colour is itself related by the Dirac quantization 
condition to the usual colour coupling $\alpha_s$ now quite well measured in 
experiment, leaving thus the mass scale of dual colour bosons as the only 
unknown parameter.  The FCNC effects so far examined include rare hadron 
decays~\cite{fcnc}, mass differences between conjugate neutral 
mesons~\cite{fcnc}, coherent muon-electron conversion on nuclei~\cite{mueconv}
and muonium conversion~\cite{mueconv}.  The conclusion is that by taking a 
mass scale for dual colour bosons of the order of 500 TeV, which value will be
seen later from an entirely different angle to have a special significance, 
all existing bounds on FCNC effects of the above 4 types are satisfied, 
although some of the bounds in the first 3 types are found to be close to 
the values predicted.  In particular, the decay 
\begin{equation}
K_L \rightarrow e^\pm \mu^\mp
\end{equation}
is estimated~\cite{fcnc} to have a branching ratio of around  $4 \times 
10^{-14}$ as compared with the present experimental bound~\cite{databook}
of $4.7 \times 10^{-12}$, while the ratio for the rate of the conversion 
reaction
\begin{equation}
\mu^- + Ti \rightarrow e^- + Ti
\end{equation} 
over the total $\mu$ absorption rate is estimated~\cite{mueconv} to be around 
$1.8 \times 10^{-12}$  as compared with the present bound~\cite{databook} 
of $4.3 \times 10^{-12}$. \footnote{note that the values given in the
references~\cite{fcnc,mueconv} were estimated with 400 TeV instead of the 
500 TeV here for the mass scale of dual colour bosons, and the rates are 
inversely proportional to the 4th power of the mass scale.}  

Of course, FCNC effects of the above types are small not because the FCNC
coupling is weak but because the mass scale of the FCNC bosons is high while 
the energy at which the FCNC processes take place is low.  At high energy 
comparable to the mass scale of the FCNC bosons, the new interaction due 
to the exchange of these bosons will become sizeable.  In other words, any 
particles carrying a generation index (dual colour), in particular neutrinos, 
will acquire thereby a strong interaction.  Although at first sight alarming, 
this prediction is in no contradiction to experiment since 500 TeV is an 
enormous energy not achievable in the laboratory whether at present or in 
the near future.  Indeed, even in astrophysics, such energies are only ever 
met with in cosmic rays.  For collisions with air nuclei, a cm energy of 500 
TeV corresponds to a primary energy of about $10^{20}$eV which has been 
observed only in about a dozen air shower events~\cite{Takeda}.  Such 
so-called post-GZK events have long been a puzzle in that they are thought 
not to be due to protons, for protons interact with the 2.7 K microwave 
background and quickly degrade from such high energies~\cite{Greisemin}.  
They could, however, be due to neutrinos if neutrinos at such energy indeed
acquire a strong interaction as predicted above.  An examination of this 
possibility reveals that some previously mysterious properties of post-GZK 
air showers could then be explained, besides giving some new predictions 
on air showers~\cite{airshowers} which can be tested by future experiments 
such as the AUGER project~\cite{Auger}.  Furthermore, this neutrino solution 
suggested for the post-GZK puzzle would give in the DSM scheme an estimate,
though a rather crude one, of around 500 TeV for the mass scale of dual 
colour bosons as the cm energy at which the interaction due to their exchange
becomes strong, and hence convert the numbers on FCNC effects given in the 
preceding paragraph into rough order-of-magnitude estimates to be tested
against experiment~\cite{fcnc,mueconv}.  However, the fact that FCNC rates 
typically depends on the 4th power of the mass of the exchanged boson 
which as explained can at present be only roughly estimated detracts from 
the stringency of these tests.  

A second area of applications which do not suffer from the noted lack of
stringency of FCNC tests arises from the fact that the DSM explanation
of fermion mass and mixing patterns relies heavily on the fact that the
fermion mass matrix rotates with changing energy scales~\cite{dsmrevs,Tsou}.  
Although a rotating mass matrix is a result of the renormalization group 
equation already in the conventional formulation of the Standard Model so long
as there is nontrivial mixing between up and down fermions~\cite{impromat}, 
the rotation is stronger in the DSM scheme since it gives the mixing itself 
as a consequence of the rotation.  Once the mass matrix rotates, then care 
has to be taken on how fermion flavour states are defined since flavour states 
defined as mass eigenstates at some scale(s) will no longer be eigenstates 
of the mass matrix at other scales.  This means that reaction amplitudes at 
an arbitrary energy scale will not be diagonal in the flavour states, leading 
thus to flavour-violations (``transmutations'') of a different nature from 
the FCNC effects considered before.  In particular, in contrast to the FCNC 
effects due to dual colour boson exchange which depends on the boson mass, 
here once the rotating fermion mass matrix is known, as it is in DSM from 
fitting the empirical mixing matrix, then transmutation effects can be
deduced as precise parameter-free predictions.  Examples of transmutation 
effects so far examined include lepton flavour-violating decays such 
as~\cite{impromat,transcay}
\begin{equation}
\psi, \Upsilon \rightarrow \mu^\pm \tau^\mp
\end{equation}  
and the photo-transmutation of leptons such as~\cite{photrans} 
\begin{equation}
\gamma e^- \rightarrow \gamma \tau^-,
\end{equation} 
with further classes of reactions under investigation.  The conclusion so 
far is that there is no violation of the existing, some already very 
stringent, experimental bounds, and that in some cases such as the 
$\mu \tau$ modes in $\psi(1S)$, $\psi(2S)$, $\Upsilon(1S)$, and 
$\Upsilon(2S)$ decay, 
for which the predicted branching ratios~\cite{transcay,impromat} being 
of the order of a few times $10^{-6}$ are within the present sensitivity 
of experiments such as BEPC, BaBar and BELLE.

Another consequence of the rotating mass matrix in the DSM scheme tied to 
its explanation of lepton mixing (neutrino oscillations) is an estimate
of the 
half-life of neutrinoless double-beta decays.   Since 
the various mass and coupling parameters involved are all known in the DSM 
scheme, a relatively precise estimate for the half-life can be made for e.g.
the decay 
\begin{equation}
{}^{76}Ge \rightarrow {}^{76}Se + 2 e^-
\end{equation}
giving a value~\cite{0nu2beta} of the order of $10^{29}$ years , i.e. some few 
orders of magnitude below the present experimental bound~\cite{Heidelberg} 
of $1.8 \times 10^{25}$ years but yet within the scope of planned future 
experiment. 

In summary, we conclude that the DSM scheme offers not only a viable solution
to the old generation puzzle, but also a host of detailed predictions all of
which have so far survived experimental tests although some of which can 
soon be subjected to close experimental scrutiny.

\end{document}